\documentclass{jfm}

\usepackage{graphicx}
\usepackage{natbib}

\usepackage{mathrsfs}

\newsavebox{\astrutbox}
\sbox{\astrutbox}{\rule[-5pt]{0pt}{20pt}}

\renewcommand{\theequation}{\arabic{section}.\arabic{equation}}

\usepackage{bm}
\usepackage{amsmath}
\usepackage{xr}

\title[Generalization of the Rotne-Prager-Yamakawa tensors
]{Generalization of the Rotne-Prager-Yamakawa mobility \\
and shear disturbance tensors}

\author[E. Wajnryb, K. A. Mizerski, P. J. Zuk and P. Szymczak]%
{
E\ls L\ls I\ls G\ls I\ls U\ls S\ls Z\ns   W\ls A\ls J\ls N\ls R\ls Y\ls B$^1$,\ns
K\ls R\ls Z\ls Y\ls S\ls Z\ls T\ls O\ls F\ns A.\ns M\ls I\ls Z\ls E\ls R\ls S\ls K\ls I$^2$\thanks{Email address for correspondence: kamiz@igf.edu.pl},\break 
P\ls A\ls W\ls E\ls L\ns J.\ns Z\ls U\ls K$^3$ \and P\ls I\ls O\ls T\ls R\ns S\ls Z\ls Y\ls M\ls C\ls Z\ls A\ls K$^3$
}

\affiliation{$^1$Department of Mechanics and Physics of Fluids, Institute of Fundamental and Technological Research, Polish Academy of Sciences, Pawinskiego 5B, 02-106, Warsaw Poland\\[\affilskip]
$^2$Department of Magnetism, Institute of Geophysics, Polish Academy of Sciences, ul. Ksiecia Janusza 64, 01-452 Warsaw, Poland\\[\affilskip]
$^3$Institute of Theoretical Physics, Faculty of Physics, University of Warsaw, Hoza 69, 00-681, Warsaw, Poland}

\pubyear{2010}
\volume{650}
\pagerange{119--126}

\begin{document}

\maketitle

\begin{abstract}
Rotne-Prager-Yamakawa approximation is one of the most commonly used methods of including hydrodynamic interactions in modelling of colloidal suspensions and polymer solutions.
The two main merits of this approximation is that it includes all long-range terms (i.e. decaying as $R^{-3}$ or slower in interparticle distances) and that the diffusion matrix is positive definite,
which is essential for Brownian dynamics modelling.
Here, we extend the Rotne-Prager-Yamakawa approach to include both translational and rotational degrees of freedom, and derive the regularizing corrections to account for overlapping particles.
Additionally, we show how the Rotne-Prager-Yamakawa approximation can be generalized for other geometries and boundary conditions.
\end{abstract}


\section{Introduction}

Particles moving in a viscous fluid induce a local flow field that affects other
particles. These long-range, many-body interactions, mediated by the solvent are commonly called 'hydrodynamic interactions' (HI).
The presence of HI is known to affect the dynamic properties of soft matter: they modify the values of diffusion coefficients in colloidal suspensions \citep{Dhont1996}, affect the characteristics of the coil-stretch transition in polymers \citep{Larson1989}, change the kinetic pathways of phase separation in binary mixtures \citep{Tanaka2001}, alter the kinetics of macromolecule adsorption on surfaces \citep{Wojtaszczyk1998} or cause the polymer migration in microchannels \citep{Usta2007}. 
They are also important in the dynamics of biological soft matter, such as DNA~\citep{Shaqfeh2005}, proteins \citep{Frembgen2009,Szymczak2011} or lipid membranes~\citep{Ando2013}. 

The proper account of hydrodynamic interactions is thus essential in simulation studies of soft matter in the flow. Unfortunately, HI depend in a complicated nonlinear way on the instantaneous positions of all particles in the system.
For a system of spheres, exact explicit expressions for the hydrodynamic interaction tensors exist in the form of the power series in interparticle distances, which may be incorporated into the simulation scheme \citep{Kim,Mazur1982,Brady-Bossis:1988,Felderhof1988,Cichocki1994}.
These are however relatively expensive numerically, thus various approximations are resorted to in order to make the computations more tractable.
The simplest one is based on the Oseen tensor, which assumes that the particles can be regarded as point force sources in the fluid. 
However, the diffusion matrix constructed in this way is not suitable for the Brownian dynamics simulations, because it becomes non-positive definite when separations between the particles become small.
This is not only unphysical (since the positivity of diffusion is a consequence of second law of thermodynamics) but also leads to numerical problems in the Brownian dynamics simulations, where a square root of diffusion matrix is needed.
Another commonly used approximation is the Rotne-Prager-Yamakawa tensor \citep{Rotne1969,Yamakawa1970}, which takes into account all the HI terms up to ${\it O}(a/r_{ij})^3$ in the expansion in the inverse distance between the particles (where $a$ is the particle radius).
Nevertheless, if the particles overlap, $r_{ij} < 2a$, the RPY tensor again looses its positive definiteness. To avoid this, a regularization for $r_{ij} < 2a$ has been proposed by \citet{Rotne1969}, which is not singular at $r_{ij}=0$ and positive definite for all the particle configurations.
The Rotne-Prager-Yamakawa tensor with this regularization is by far the most popular method of accounting for HI in soft matter modelling \citep{Naegele2006}. Notably, RPY tensor is divergence-free, which considerably simplifies the application of Brownian dynamics algorithm \cite{Ermak1978}. However, when one goes beyond RPY approximation and includes many-body effects in hydrodynamic interactions, the divergence of mobility matrix becomes is non-zero and needs to be taken into account in Brownian dynamics simulation schemes \citep{Wajnryb2004}.

The present paper takes a close look at the Rotne-Prager-Yamakawa approximation and generalizes it in a number of ways.
First, we re-derive the original RPY tensor using the direct integration of force densities over the sphere surfaces. When the spheres overlap then this method gives us automatically the regularization correction.
In this way we derive the RPY regularizations not only for the translational degrees of freedom (already obtained by \citet{Rotne1969}) but also for rotational degrees of freedom, as well as for the shear disturbance matrix ${\bf{C}}$ - another hydrodynamic tensor, which gives the response of the particles to the external shear flow.
The mobility evaluated using our technique may be applied for calculation of the diffusion tensor
of complex molecules \citep{Adamczyk2012,Garcia2007} using bead models which include overlapping spheres.
Finally, we show how these results can be generalized for other boundary conditions and corresponding propagators.

\section{The mobility problem under shear flow}

We consider a suspension of $N$ identical spherical particles of radius
$a$, in an incompressible fluid of viscosity $\eta$ at a low Reynolds number. 
The particles are immersed in a linear shear flow
\begin{equation}
\mathbf{v}_{\infty}\left(\mathbf{r}\right)=\mathbf{K}_{\infty}\bm{\cdot}\mathbf{r},\label{eq:Shear_flow}
\end{equation}
where $\mathbf{K}_{\infty}$ is the constant velocity gradient matrix, e.g. for a simple shear flow
\begin{equation}
\mathbf{K}_{\infty}=\left[\begin{array}{ccc}
0 & 0 & \dot{\gamma}\\
0 & 0 & 0\\
0 & 0 & 0
\end{array}\right],\qquad\dot{\gamma}=\mathrm{{const.}}\label{eq:shear_matrix}
\end{equation}

Due to the linearity of the Stokes equations, the forces and torques exerted by the fluid on the particles ($\bm{\mathcal{F}}_j$ and $\bm{\mathcal{T}}_j$)  depend linearly on translational and rotational velocities of the particles ($\mathbf{U}_i$, $\bm{\Omega}_{i}$). This relation defines the generalized friction matrix $\bm{\zeta}$
\begin{equation}
\left( \begin{array}{c}
\bm{\mathcal{F}}_j  \\
\bm{\mathcal{T}}_j  \\
\end{array} \right)
= - \sum_{i}
\left( \begin{array}{ccc}
\bm{\zeta}^{tt}_{ji} & \bm{\zeta}^{tr}_{ji} & \bm{\zeta}^{td}_{ji} \\
\bm{\zeta}^{rt}_{ji} & \bm{\zeta}^{rr}_{ji} & \bm{\zeta}^{rd}_{ji} 
\end{array} \right)
\cdot
\left( \begin{array}{c}
\mathbf{v}_{\infty}({\bf R}_{i}) - \mathbf{U}_{i} \\
\bm{\omega}_{\infty}({\bf R}_{i}) - \bm{\Omega}_{i} \\
\mathbf{E}_{\infty}
\end{array} \right),
\label{eq:generalFrictionMatrix}
\end{equation}
where $ \bm{\zeta}^{pq}$ (with $p=t,r$ and $q=t,r,d$) are the Cartesian tensors and the superscripts $t$, $r$ and $d$ correspond to the translational, rotational and dipolar components, respectively. The tensor $\mathbf{E}_{\infty}$ is the symmetric part of $\mathbf{K}_{\infty}$ in (\ref{eq:shear_matrix}) and
$\bm{\omega}_{\infty} = \frac{1}{2} \nabla \times \mathbf{v}_{\infty} ({\bf R}_i) =\frac{1}{2} \bm{\epsilon}:\mathbf{K}_{\infty}$ is the vorticity of the incident flow. Finally $\mathbf{R}_i$ corresponds to  the position of particle $i$. The reciprocal relation giving velocities of particles moving under external forces/torques in external flow $\mathbf{v}_{\infty}$
is determined by generalized mobility matrix $\bm{\mu}$ written after \citet{Dhont1996}
\begin{equation}
\left( \begin{array}{c}
\mathbf{U}_{i} \\
\bm{\Omega}_{i}
\end{array} \right)
= 
\left( \begin{array}{c}
\mathbf{v}_{\infty}({\bf R}_i)  \\
\bm{\omega}_{\infty}({\bf R}_i)
\end{array} \right)
+
\sum_{j}
\left[
\left( \begin{array}{cc}
\bm{\mu}^{tt}_{ij} & \bm{\mu}^{tr}_{ij}  \\
\bm{\mu}^{rt}_{ij} & \bm{\mu}^{rr}_{ij} 
\end{array} \right)
\cdot
\left( \begin{array}{c}
\bm{\mathcal{F}}_{j}  \\
\bm{\mathcal{T}}_{j}
\end{array} \right)
\right]
+
\left( \begin{array}{c}
\mathbf{C}^{t}_{i}  \\
\mathbf{C}^{r}_{i} 
\end{array} \right)
:
\mathbf{E}_{\infty},
\label{eq:generalMobilityMatrixSeparatedC}
\end{equation}
where the shear disturbance tensor $\mathbf{C}$ elements are defined as
\begin{subequations} \label{eq:ctensor}
\begin{equation}
\mathbf{C}^{t}_{i}  = \sum_{j} \bm{\mu}^{td}_{ij}, \quad
\mathbf{C}^{r}_{i}  = \sum_{j} \bm{\mu}^{rd}_{ij}.
\end{equation}
\end{subequations}
In the case of single particle the mobility matrixes reduce to
\begin{equation} \label{eq:isolatedParticles}
 \bm{\mu}^{tt}_{ii} = \frac{1}{\zeta^{tt}} {\bf 1}, \quad \bm{\mu}^{rr}_{ii} = \frac{1}{\zeta^{rr}} {\bf 1}, \quad \bm{\mu}^{tr}_{ii} = \bm{\mu}^{rt}_{ii} = 0,
\end{equation}
where the friction coefficients for a spherical particle are given by $\zeta^{tt}=6\pi\eta a$ and $\zeta^{rr}=8\pi\eta a^{3}$.

Finding the mobility matrix (or the associated diffusion matrix, ${\bf D}= k_B T \bm{\mu}$) is the problem of a fundamental importance in constructing the numerical algorithms for tracking the motion of the particles in viscous fluid. The two main numerical methods used for this purpose are the {\it Stokesian Dynamics},
which corresponds to the numerical integration of~\ref{eq:generalMobilityMatrixSeparatedC} and the {\it Brownian dynamics}, used whenever the Brownian motion of the particles cannot be neglected~\citep{Naegele2006}.
In the latter, the random displacements of the particles, $\bm{\Gamma}_i(\Delta t)$, need to be added on top of the deterministic displacements governed by Eq.~\ref{eq:generalMobilityMatrixSeparatedC}.
The fluctuation-dissipation theorem implies that the covariance of $\bm{\Gamma}$ is connected to the mobility matrix, e.g. for the translational displacements 
\begin{equation}
\left\langle \bm{\Gamma}_i\left(\bigtriangleup t\right)\bm{\Gamma}_j\left(\bigtriangleup t\right)\right\rangle =2k_{B}T\bm{\mu}^{tt}_{ij} \bigtriangleup t. \label{eq:Fluct_diss}
\end{equation}
Hence the calculation of $\bm{\Gamma}_i(\Delta t)$ requires finding a matrix ${\bf d}$ such that $\bm{\mu}^{tt} = {\bf d} {\bf d}^T$.
This is possible only when the mobility matrix is positively defined.
Any valid approximation scheme for the hydrodynamic interactions should then not only correctly reproduce the particle mobilities but also guarantee the positive definiteness of the mobility tensors. 

\section{The Rotne-Prager-Yamakawa form of $\bm{\mu}$ and $\mathbf{C}$ for
systems with shear.}

In principle the hydrodynamic interactions tensors can be calculated with arbitrary precision, following e.g. the multipole expansion or boundary integral method \citep{Kim,Pozrikidis1992}. In practice, however, the exact approach turns out to be too demanding computationally,  so various approximation procedures have to be resorted to. The most commonly used is the Rotne-Prager-Yamakwa approximation \citep{Rotne1969,Yamakawa1970},
based on the following idea: when a force (or torque) is applied to particle $i$, that particle begins to move inducing the flow in the bulk of the fluid.
The extent to which this additional flow affects translational and rotational velocities of another particle ($j$) is then calculated using Faxen's laws \citep{Kim}. In that way one neglects not only the multi-body effects (involving three and more particles) but also the higher order terms in two-particle interactions (e.g. we do not consider the impact of the movement of particle $j$ back on particle $i$). Below, we follow this procedure to derive in a systematic way 
hydrodynamic tensors for both translational and rotational degrees of freedom.

\subsection{The mobility matrix $\boldsymbol{\mu}$}

The Stokes flow generated by a point force in the unbounded space is given by the Oseen Tensor \citep{Kim}
\begin{equation}
\mathbf{T}_{0}\left(\mathbf{r}\right)=\frac{1}{8\pi\eta r}\left(\mathbf{1}+\hat{\mathbf{r}}\hat{\mathbf{r}}\right).\label{eq:Oseen_def}
\end{equation}
Since $\mathbf{T}_{0}\left(\mathbf{r}\right)$  is a Green function for Stokes equations, one can use it to calculate the translational $\mathbf{v}^t_0(\mathbf{r})$ and rotational $\mathbf{v}^r_0(\mathbf{r})$ flows generated by the sphere situated~at~$\mathbf{R}_{j}$,
to which we apply  force $\bm{\mathcal{F}}$ and/or torque $\bm{\mathcal{T}}$: 
\begin{eqnarray}
\mathbf{v}_{0}^{t}\left(\mathbf{r}\right) & = & \int_{S_{j}}\mathbf{T}_{0}\left(\mathbf{r}-\mathbf{r}'\right)\cdot\frac{\bm{\mathcal{F}}}{4\pi a^{2}}\mathrm{d}\sigma'\nonumber \\
 & = & \left\{ \begin{array}{c}
\left(1+\frac{a^{2}}{6}\nabla^{2}\right)\mathbf{T}_{0}\left(\bm{\rho}_j \right)\cdot\bm{\mathcal{F}}=\frac{1}{8\pi\eta \rho_j}\left[\left(1+\frac{a^{2}}{3 \rho_j^{2}}\right)\mathbf{1}+\left(1-\frac{a^{2}}{\rho_j^{2}}\right)\hat{\bm{\rho}}_j\hat{\bm{\rho}}_j\right]\cdot\bm{\mathcal{F}},\qquad \rho_j>a,\\
\,\\
\frac{1}{\zeta^{tt}}\bm{\mathcal{F}},\qquad \rho_j \leq a,
\end{array}\right.\label{eq:v0t}
\end{eqnarray}
\begin{equation}
\mathbf{v}_{0}^{r}\left(\mathbf{r}\right)=\int_{S_{j}}\mathbf{T}_{0}\left(\mathbf{r}-\mathbf{r}'\right)\cdot\frac{3}{8\pi a^{3}}\bm{\mathcal{T}}\times \mathbf{n}' \mathrm{d}\sigma'=
\left\{ \begin{array}{c}
\frac{1}{2}\nabla\times\mathbf{T}_{0}\left(\bm{\rho}_j\right)\cdot\bm{\mathcal{T}}=\frac{1}{8\pi\eta \rho_j^{3}} \bm{\mathcal{T}} \times \bm{\rho}_j,\quad \rho_j>a,\\
\\
\frac{1}{\zeta^{rr}} \bm{\mathcal{T}} \times \bm{\rho}_j  ,\qquad \rho_j \leq a,
\end{array}\right.\label{eq:v0r}
\end{equation}
\noindent where $\bm{\rho}_j = \mathbf{r} - \mathbf{R}_j$ is the distance from the sphere centre,  $\mathbf{r}'$ denotes integration variable,
$\mathbf{n}'$ is the unit normal vector to the sphere at point $\mathbf{r}'$ and $\int_{S_{j}}$ denotes an integral over the surface
of the sphere situated at $\mathbf{R}_j$. The curl of a tensor is defined in the following way
\begin{equation}
\left(\nabla\times\mathbf{T}\right)_{\alpha\beta}=\epsilon_{\alpha\gamma\zeta}\partial_{\gamma}\mathbf{T}_{\zeta\beta}.\label{eq:curl_of_a_tensor-1}
\end{equation}
where the Greek letters denote the Cartesian components.

The Faxen laws \citep{Kim} allow to express the velocity
$\mathbf{U}_i$ and angular velocity $\bm{\Omega}_i$ of a sphere $i$ immersed
in an external flow $\mathbf{v}_{0}$ placed in $\mathbf{R}_i$
\begin{equation}
\mathbf{U}_i=\frac{1}{4\pi a^{2}}\int_{S_i}\mathbf{v}_{0}\left(\mathbf{r}'\right)\mathrm{d}\sigma'=\left.\left(\mathbf{1}+\frac{a^{2}}{6}\nabla^{2}\right)\mathbf{v}_{\mathrm{0}} \right|_{\mathbf{r}=\mathbf{R}_i}, \nonumber
\end{equation}
\begin{equation}
\bm{\Omega}_i=\frac{3}{8\pi a^{3}}\int_{S_i}\mathbf{n}'\times\mathbf{v}_{0}\left(\mathbf{r}'\right)\mathrm{d}\sigma'=\left. \frac{1}{2}\nabla\times\mathbf{v}_{0} \right|_{\mathbf{r}=\mathbf{R}_i} ,\label{eq:Faxen_Laws}
\end{equation}
\noindent where the integration is performed over the sphere surface $S_i$.
Thus substituting (\ref{eq:v0t},\ref{eq:v0r}) into equations (\ref{eq:Faxen_Laws}) we obtain the contribution to velocity $\mathbf{U}'_i$ and angular velocity $\bm{\Omega}'_i$
of a sphere $i$ due to the force/torque acting on a sphere $j$
\begin{eqnarray}
\mathbf{U}'_i & = & \frac{1}{4\pi a^{2}}\int_{S_{i}}\mathbf{v}_{0}^{t}\left(\mathbf{r}'\right)\mathrm{d}\sigma'+\frac{1}{4\pi a^{2}}\int_{S_{i}}\mathbf{v}_{0}^{r}\left(\mathbf{r}'\right)\mathrm{d}\sigma' \nonumber \\
 & = & \frac{1}{4\pi a^{2}}\int_{S_{i}}\mathrm{d}\sigma'\int_{S_{j}}\mathrm{d}\sigma''\mathbf{T}_{0}\left(\mathbf{r}'-\mathbf{r}''\right)\cdot\left[\frac{\bm{\mathcal{F}}}{4\pi a^{2}}+\frac{3}{8\pi a^{3}}\bm{\mathcal{T}}\times \mathbf{n}''\right],\label{eq:U_final}
\end{eqnarray}
\begin{eqnarray}
\bm{\Omega}'_i & = & \frac{3}{8\pi a^{3}}\int_{S_{i}}\mathbf{n}' \times\mathbf{v}_{0}^{t}\left(\mathbf{r}'\right)\mathrm{d}\sigma'+\frac{3}{8\pi a^{3}}\int_{S_{i}}\mathbf{n}'\times\mathbf{v}_{0}^{r}\left(\mathbf{r}'\right)\mathrm{d}\sigma'\nonumber \\
 & = & \frac{3}{8\pi a^{3}}\int_{S_{i}}\mathrm{d}\sigma'\int_{S_{j}}\mathrm{d}\sigma''\mathbf{n}'\times\mathbf{T}_{0}\left(\mathbf{r}'-\mathbf{r}''\right)\cdot\left[\frac{\bm{\mathcal{F}}}{4\pi a^{2}}+\frac{3}{8\pi a^{3}}\bm{\mathcal{T}}\times \mathbf{n}''\right].\label{eq:Omega_final}
\end{eqnarray}
At this stage let us introduce tensors
\begin{equation}
\mathbf{w}_i^{t}\left(\mathbf{r}\right)=\frac{1}{4\pi a^{2}}\mathbf{1}\delta\left( \rho_j - a\right),\qquad\mathbf{w}_i^{r}\left(\mathbf{r}\right)=\frac{3}{8\pi a^{3}}\bm{\epsilon}\cdot\hat{\bm{\rho}}_j\delta\left( \rho_j - a\right),\label{eq:wtr}
\end{equation}
\noindent where $(\boldsymbol{\epsilon}\cdot\hat{\bm{\rho}}_j)_{\alpha\beta}=\epsilon_{\alpha\beta\gamma}\hat{\rho}_{j \gamma}$.
Above tensors multiplied by force $\mathbf{w}^t \cdot \bm{\mathcal{F}}$ and torque $\mathbf{w}^r \cdot \bm{\mathcal{T}}$
have the interpretation of the force densities on the surface of the sphere due to the force and torque acting on the sphere.
We can now write down the following general formulae for the mobility matrix
\begin{equation}
\bm{\mu}_{ij}^{tt}=\left\langle \mathbf{w}_{i}^{t}\right|\mathbf{T}_{0}\left|\mathbf{w}_{j}^{t}\right\rangle ,\quad \bm{\mu}_{ij}^{rr}=\left\langle \mathbf{w}_{i}^{r}\right|\mathbf{T}_{0}\left|\mathbf{w}_{j}^{r}\right\rangle ,\quad\bm{\mu}_{ij}^{rt}=\left\langle \mathbf{w}_{i}^{r}\right|\mathbf{T}_{0}\left|\mathbf{w}_{j}^{t}\right\rangle, \quad \bm{\mu}_{ij}^{tr}=\left\langle \mathbf{w}_{i}^{t}\right|\mathbf{T}_{0}\left|\mathbf{w}_{j}^{r}\right\rangle ,\label{eq:mu_formulae_gen}
\end{equation}
where we use the bra-ket notation defined in the following way
\begin{equation}
\bm{\mu}_{ij}^{pq}=\left\langle \mathbf{w}_{i}^{p}\right|\mathbf{T}_{0}\left|\mathbf{w}_{j}^{q}\right\rangle =\int\mathrm{d}\mathbf{r}'\int\mathrm{d}\mathbf{r}''\left[\mathbf{w}_{i}^{p}\left(\mathbf{r}'\right)\right]^{T}\cdot\mathbf{T}_{0}\left(\mathbf{r}'-\mathbf{r}''\right)\cdot\mathbf{w}_{j}^{q}\left(\mathbf{r}''\right),\label{eq:Ypq}
\end{equation}
with $p,q=r,t$ and $T$ - tensor transposition. The method of calculation of the integrals
in (\ref{eq:U_final})-(\ref{eq:Omega_final}) is presented in the Supplementary Material.
Here, we simply quote the final results denoting $\mathbf{R}_{ij}= \mathbf{R}_{i} - \mathbf{R}_j$.
Fortunately there is no need to integrate explicitly for non overlapping spheres. For the translational-translational mobility, we get:
\begin{equation}
\bm{\mu}_{ij}^{tt}=\left\{ \begin{array}{c}
\left(\mathbf{1}+\frac{a^{2}}{3}\nabla^{2}\right)\mathbf{T}_{0}\left(\mathbf{R}_{ij}\right)=\frac{1}{8\pi\eta R_{ij}}\left[\left(1+\frac{2a^{2}}{3R_{ij}^{2}}\right)\mathbf{1}+\left(1-\frac{2a^{2}}{R_{ij}^{2}}\right)\hat{\mathbf{R}}_{ij}\hat{\mathbf{R}}_{ij}\right],\quad R_{ij}>2a,\\
\\
\frac{1}{\zeta^{tt}}\left[\left(1-\frac{9R_{ij}}{32a}\right)\mathbf{1}+\frac{3R_{ij}}{32a}\hat{\mathbf{R}}_{ij}\hat{\mathbf{R}}_{ij}\right],\qquad R_{ij}\leq2a,
\end{array}\right.\label{eq:mutt_final}
\end{equation}
which, in the limit of $R_{ij} \rightarrow 0$, yields the self mobility
\begin{equation}
\bm{\mu}^{tt}_{ii}=\bm{\mu}^{tt}_{jj}= \underset{R_{ij}\rightarrow0}{\mathrm{lim}}\,\bm{\mu}^{tt}_{ij} = \frac{1}{\zeta^{tt}}\mathbf{1}.\label{eq:self_tt}
\end{equation}
Next, for the rotational degrees of freedom
\begin{equation}
\bm{\mu}_{ij}^{rr}=\left\{ \begin{array}{c}
-\frac{1}{4}\nabla^{2}\mathbf{T}_{\mathrm{0}}\left(\mathbf{R}_{ij}\right)=-\frac{1}{16\pi\eta R_{ij}^{3}}\left(\mathbf{1}-3\hat{\mathbf{R}}_{ij}\hat{\mathbf{R}}_{ij}\right),\qquad R_{ij}>2a,\\
\\
\frac{1}{\zeta^{rr}}\left[\left(1-\frac{27}{32}\frac{R_{ij}}{a}+\frac{5}{64}\frac{R_{ij}^{3}}{a^{3}}\right)\mathbf{1}+\left(\frac{9}{32}\frac{R_{ij}}{a}-\frac{3}{64}\frac{R_{ij}^{3}}{a^{3}}\right)\hat{\mathbf{R}}_{ij}\hat{\mathbf{R}}_{ij}\right],\qquad R_{ij}\leq2a,
\end{array}\right.\label{eq:murr_final}
\end{equation}
with the self mobility given by
\begin{equation}
\bm{\mu}^{rr}_{ii}=\bm{\mu}^{rr}_{jj}=\underset{R_{ij}\rightarrow0}{\mathrm{lim}}\,\bm{\mu}^{rr}_{ij}=\frac{1}{\zeta^{rr}}\mathbf{1}.
\end{equation}
Finally, the translational-rotational mobility is described by the following tensor
\begin{equation}
\bm{\mu}_{ij}^{rt} = \left[\bm{\mu}_{ij}^{tr}\right]^{T}=\left\{ \begin{array}{c}
\frac{1}{2}\nabla\times\left(\mathbf{1}+\frac{a^{2}}{6}\nabla^{2}\right)\mathbf{T}_{\mathrm{0}}\left(\mathbf{R}_{ij}\right)=\frac{1}{2}\nabla\times\mathbf{T}_{0}\left(\mathbf{R}_{ij}\right)=\frac{1}{8\pi\eta R_{ij}^{2}}\boldsymbol{\epsilon}\cdot\hat{\mathbf{R}}_{ij},\qquad R_{ij}>2a,\\
\\
\frac{1}{16\pi\eta a^{2}}\left(\frac{R_{ij}}{a}-\frac{3}{8}\frac{R_{ij}^{2}}{a^{2}}\right)\boldsymbol{\epsilon}\cdot\hat{\mathbf{R}}_{ij},\qquad R_{ij}\leq2a,
\end{array}\right.\label{eq:murt_final}
\end{equation}
with
\begin{equation}\label{eq:mutr_final_self}
\bm{\mu}^{tr}_{ii}=\bm{\mu}^{tr}_{jj}=\bm{\mu}^{rt}_{ii}=\bm{\mu}^{rt}_{jj}=\underset{R_{ij}\rightarrow0}{\mathrm{lim}}\,\bm{\mu}^{tr}_{ij}=\underset{R_{ij}\rightarrow0}{\mathrm{lim}}\,\bm{\mu}^{rt}_{ij}= \mathbf{0}.
\end{equation}
Note that the formulae \eqref{eq:mutt_final} for the translational  mobility matrix, both for $R_{ij}>2a$ and for $R_{ij}<2a$ were derived earlier by \citet{Rotne1969} and \citet{Yamakawa1970} and are known as Rotne-Prager-Yamakawa mobility approximation.
The expressions for the other components of the mobility matrix $\bm{\mu}_{ij}^{rr}$ and $\bm{\mu}_{ij}^{tr}$ are also known \citep{Kim,Dhont1996,Reichert2006,Garcia2007} but only for $R_{ij}>2a$.
However, to our knowledge, the regularizing corrections for $\bm{\mu}_{ij}^{rr}$ and $\bm{\mu}_{ij}^{tr}$ for the overlapping particles ($R_{ij}<2a$) have not been derived so far.
Importantly, as we will demonstrate in the section (\ref{sec:pos_def}), only with the use of these corrections the mobility matrix $\bm{\mu}$ remains positive definite for all configurations of the particles. 

Contrastingly, in the point-force (Stokeslet) model which is sometimes used for modelling the dynamics of colloidal suspensions \citep{Pear1981}, the mobility matrix, defined as follows
\begin{equation}\label{eq:point_force}
\bm{\mu}^{tt}_{ij}=\frac{1}{8\pi\eta R_{ij}}\left(\mathbf{1}+\hat{\mathbf{R}}_{ij}\hat{\mathbf{R}}_{ij}\right), \quad i \neq j,
\qquad \bm{\mu}^{tt}_{ii}=\frac{1}{\zeta^{tt}} {\bf 1},
\end{equation}
is not positive definite even for non overlapping spheres and does not possess the property (\ref{eq:self_tt}).

\subsection{The shear disturbance tensor $\mathbf{C}$}

The formula for the 3rd rank convection tensor $\mathbf{C}$ can be obtained in the following way.
\citet{Kim} provide a solution for the excess flow $\mathbf{v}^{c}_{0}\left(\mathbf{r}\right)$,
produced by a freely moving sphere situated at $\mathbf{R}_j$ in the ambient shear flow $\mathbf{K}_{\infty}\cdot\mathbf{r}$,
which is a difference between total flow $\mathbf{v} \left(\mathbf{r}\right)$ and ambient flow
\begin{equation}
\mathbf{v}^{c}_{0} \left(\mathbf{r}\right) = \mathbf{v}\left(\mathbf{r}\right)-\mathbf{K}_{\infty}\cdot\mathbf{r}= \frac{20}{3} \pi \eta a^3 \left\{ \left[\left(1+\frac{a^{2}}{10}\nabla^{2}\right)\mathbf{T}_{0}\left(\bm{\rho}_j\right)\right]\overleftarrow{\bm{\nabla}}\right\} :\mathbf{E}_{\infty},\label{eq:KimKarrila_flow-1}
\end{equation}
where $\left[\mathbf{T}(\mathbf{r})\overleftarrow{\bm{\nabla}}\right]_{\alpha\beta\gamma}=\partial_{\gamma}T_{\alpha\beta}(\mathbf{r})$.
The contribution to the surface force density due to the straining fluid motion is: $3\eta \delta\left( \rho_j - a \right)\mathbf{E}_{\infty}\cdot\hat{\bm{\rho}_j}$, thus introducing tensor $\mathbf{w}^{c}\left(\mathbf{r}\right)$
\begin{equation}
\mathbf{w}^{c}\left(\mathbf{r}\right) :\mathbf{E}_{\infty} = 3\eta \delta\left( \rho_j - a \right)\mathbf{E}_{\infty}\cdot\hat{\bm{\rho}_j},\label{eq:wc}
\end{equation}
and using the Green's formula we may express the excess
flow over the shear flow $\mathbf{K}_{\infty}\cdot\mathbf{r}$ in the following
way
\begin{equation}
\mathbf{v}_{0}^{c}\left(\mathbf{r}\right)=3\eta \int_{S_{j}}\mathbf{T}_{0}\left(\mathbf{r}-\mathbf{r}'\right)\cdot\mathbf{E}_{\infty}\cdot \mathbf{n}'\mathrm{d}\sigma'=\left\{ \begin{array}{c}
\frac{20}{3} \pi \eta a^3 \left\{ \left[\left(1+\frac{a^{2}}{10}\nabla^{2}\right)\mathbf{T}_{\mathrm{0}}\left(\bm{\rho}_j\right)\right]\overleftarrow{\bm{\nabla}}\right\} :\mathbf{E}_{\infty},\quad \rho_j >a,\\
\\
-\mathbf{E}_{\infty}\cdot\bm{\rho}_j,\qquad \rho_j \leq a.
\end{array}\right.\label{eq:vc}
\end{equation}
Now, by the Faxen laws (\ref{eq:Faxen_Laws}) the contribution to velocity
and angular velocity of another sphere (say number $i$) immersed in such flow is
\begin{equation}
\mathbf{U}'_i=\left\langle \mathbf{w}_{i}^{t}\right|\mathbf{T}_{0}\left|\mathbf{w}_{j}^{c}\right\rangle :\mathbf{E}_{\infty}=\frac{1}{4\pi a^{2}}\int_{S_{i}}\mathbf{v}_{0}^{c}\left(\mathbf{r}'\right)\mathrm{d}\sigma'=\bm{\mu}^{td}_{ij}:\mathbf{E}_{\infty},\label{eq:u_C_integral}
\end{equation}
\begin{equation}
\bm{\Omega}'_i=\left\langle \mathbf{w}_{i}^{r}\right|\mathbf{T}_{0}\left|\mathbf{w}_{j}^{c}\right\rangle :\mathbf{E}_{\infty}=\frac{3}{8\pi a^{3}}\int_{S_{i}} \mathbf{n}' \times \mathbf{v}_{0}^{c}\left(\mathbf{r}'\right)\mathrm{d}\sigma'=\bm{\mu}^{rd}_{ij}:\mathbf{E}_{\infty}.\label{eq:Omega_C_integral}
\end{equation}
For the case of $R_{ij}>2a$, the form of $\bm{\mu}^{td}_{ij}$, $\bm{\mu}^{rd}_{ij}$ is expressed using (\ref{eq:Faxen_Laws}) and (\ref{eq:vc}) in terms
of differential operators
\begin{equation}
\bm{\mu}^{td}_{ij}:\mathbf{E}_{\infty}=\frac{20}{3} \pi \eta a^3 \left\{ \left[ \left( 1 + \frac{4a^{2}}{15}\nabla^{2} \right) \mathbf{T}_{\mathrm{0}}\left(\mathbf{R}_{ij}\right)\right]\overleftarrow{\bm{\nabla}}\right\}:\mathbf{E}_{\infty},\label{eq:C_Hasimoto_def-1}
\end{equation}
\begin{equation}
\bm{\mu}^{rd}_{ij}:\mathbf{E}_{\infty}=\frac{10}{3} \pi \eta a^3 \left\{\left[\nabla\times\mathbf{T}_{\mathrm{0}}\left(\mathbf{R}_{ij}\right)\right]\overleftarrow{\bm{\nabla}}\right\}:\mathbf{E}_{\infty},\label{eq:C_r_definition-1}
\end{equation}
\noindent where $\bm{\nabla}$ denotes derivation with respect to
$R_{ij}^{\alpha}$ and $\left[\mathbf{T}(\mathbf{R}_{ij})\overleftarrow{\bm{\nabla}}\right]_{\alpha\beta\gamma}=\partial_{\gamma}T_{\alpha\beta}(\mathbf{R}_{ij})$.
This allows to write down the final results in the following form
\begin{equation}
\left[ \mu^{td}_{ij} \right] _{\alpha\beta\gamma}=\left\{ \begin{array}{c}
\frac{5}{6} a \left[-\frac{16}{5}\frac{a^{4}}{R_{ij}^{4}}\hat{R}_{ij}^{\gamma}\delta_{\alpha\beta}+\left(-3\frac{a^{2}}{R_{ij}^{2}}+8\frac{a^{4}}{R_{ij}^{4}}\right)\hat{R}_{ij}^{\alpha}\hat{R}_{ij}^{\beta}\hat{R}_{ij}^{\gamma}\right],\qquad R_{ij}>2a,\\
\\
\frac{5}{6} a \left[\left(-\frac{3}{5}\frac{R_{ij}}{a}+\frac{1}{4}\frac{R_{ij}^{2}}{a^{2}}\right)\hat{R}_{ij}^{\gamma}\delta_{\alpha\beta}-\frac{1}{16}\frac{R_{ij}^{2}}{a^{2}}\hat{R}_{ij}^{\alpha}\hat{R}_{ij}^{\beta}\hat{R}_{ij}^{\gamma}\right],\qquad R_{ij}\leq2a,
\end{array}\right.\label{eq:ct}
\end{equation}
with the respective limit in the self case
\begin{equation}
\bm{\mu}^{td}_{ii}=\bm{\mu}^{td}_{jj}=\underset{R_{ij}\rightarrow0}{\mathrm{lim}}\,\bm{\mu}^{td}_{ij}=\mathbf{0},
\end{equation}
\begin{equation}
\left[ \mu^{rd}_{ij} \right] _{\alpha\beta\gamma}=\left\{ \begin{array}{c}
-\frac{5}{2}\left(\frac{a}{R_{ij}}\right)^{3}\epsilon_{\alpha\beta\zeta}\hat{R}_{ij}^{\zeta}\hat{R}_{ij}^{\gamma},\qquad R_{ij}>2a,\\
\\
-\frac{5}{2} \left(\frac{3}{16}\frac{R_{ij}}{a}-\frac{1}{32}\frac{R_{ij}^{3}}{a^{3}}\right)\epsilon_{\alpha\beta\zeta}\hat{R}_{ij}^{\zeta}\hat{R}_{ij}^{\gamma},\qquad R_{ij}\leq2a,
\end{array}\right.\label{eq:cr}
\end{equation}
and in the self case as limit
\begin{equation} \label{eq:mu_rdself}
\bm{\mu}^{rd}_{ii}=\bm{\mu}^{rd}_{jj}=\underset{R_{ij}\rightarrow0}{\mathrm{lim}}\,\bm{\mu}^{rd}_{ij}=\mathbf{0}.
\end{equation}
The expressions for $R_{ij}<2a$ in (\ref{eq:ct}) and (\ref{eq:cr}) vanish for $\mathbf{R}_{ij}=0$
and match with the $R_{ij}>2a$ expressions at $R_{ij}=2a$.

Note that (\ref{eq:C_Hasimoto_def-1}) and (\ref{eq:C_r_definition-1}) do not determine $\bm{\mu}^{td}$ and $\bm{\mu}^{rd}$ uniquely,
since they define only the symmetric and traceless parts of mobility matrix.
Given this freedom, in (\ref{eq:ct})-(\ref{eq:mu_rdself}) we take the matrices in the simplest algebraic form.

This completes our derivation making all the terms in mobility equation (\ref{eq:generalMobilityMatrixSeparatedC})
directly computable under the Rotne-Prager-Yamakawa approximation.

\subsection{Positive definiteness}
\label{sec:pos_def}

It is now a straightforward task to demonstrate the positive definiteness of the mobility matrix given by (\ref{eq:Ypq}). 
\citet{Cichocki2000} provide a simple proof of positive definiteness of a quadratic form such as in (\ref{eq:Ypq}), which we will now
summarize. Consider the following quadratic form
\begin{equation}
\left\langle \mathbf{g}\right|\mathbf{T}_{0}\left|\mathbf{g}\right\rangle =\int\mathrm{d}\mathbf{r}\int\mathrm{d}\bar{\mathbf{r}}\mathbf{g}\left(\mathbf{r}\right)^{*}\cdot\mathbf{T}_{0}\left(\mathbf{r}-\bar{\mathbf{r}}\right)\cdot\mathbf{g}\left(\bar{\mathbf{r}}\right) ,\label{eq:PD_1}
\end{equation}
where $\mathbf{g}(\mathbf{r})$ is a complex valued function and the upper star is complex conjugation.
We will show that from positive definiteness of $\mathbf{T}_0$ follows that $\bm{\mu}^{pq}$ is positive definite. Let 
\begin{equation}
\mathbf{d}\left(\mathbf{r}\right)=\sum_{i,p} \mathbf{w}_{i}^{p}\left(\mathbf{r}\right) \cdot \mathbf{d}^{p}_{i}, \label{eq:PDz_2}
\end{equation}
where $\mathbf{d}^{p}_{i}$ denotes an arbitrary vector. Now we write
\begin{equation}
0 \leq \left\langle \mathbf{d} | \mathbf{T}_0 | \mathbf{d} \right\rangle = \sum_{i,p} \sum_{j,q} {\mathbf{d}^{p}_{i}}^* \cdot \bm{\mu}^{pq}_{ij} \cdot \mathbf{d}^{q}_{j} \label{eq:PDz_3},
\end{equation}
which ends the proof.

Note that the above proof of positivity does not hold for the point-force model (\ref{eq:point_force}).
In this case the off-diagonal ($i\neq j$) terms of the mobility matrix can be cast in the form (\ref{eq:Ypq}) 
using $\mathbf{w}_i^{t}(\mathbf{r})=\mathbf{1}\delta(\mathbf{r}-\mathbf{R}_i)$. The diagonal terms, however, would then become infinite due to the singularity at $R_{ij}=0$.
This problem is circumvented in the formulation (\ref{eq:point_force}) by using single-particle mobilities 
$1/\zeta^{tt}$ for the diagonal terms. However, the resulting point-force mobility matrix is not positive definite for arbitrary configuration,
thus cannot be used in Brownian dynamics simulations.

\section{Generalization of the Rotne-Prager-Yamakawa mobility for arbitrary propagator}

In this section we consider a general case of particles interacting
hydrodynamically e.g. in confined geometry, periodic boundary
conditions or in the presence of interfaces. We assume that for a
given geometry a positive-definite Green's function,
$\mathbf{T}(\mathbf{r},\mathbf{r}')$,  can be derived.  Such solutions
have indeed been constructed, e.g. for systems bounded by a cylinder
and a sphere~\citep{Lorentz,Liron,Oseen}, for periodic system
\citep{Hasimoto} as well as for the system bounded by one \citep{Blake}
and two walls \citep{Bhattacharya2005}.

We define the Rotne-Prager-Yamakawa approximation for the positive definite mobility matrix in analogous way to (\ref{eq:Ypq})
\begin{equation}
\bm{\mu}_{ij}^{pq}=\left\langle \mathbf{w}_{i}^{p}|\mathbf{T}|\mathbf{w}_{j}^{q}\right\rangle =\int \int d\mathbf{r'} ~ d\mathbf{r''~[w}_{i}^{p}(\mathbf{r'})]^{\mathrm{T}}\cdot \mathbf{T(r',r''})\cdot \mathbf{w}_{j}^{q}(\mathbf{r''}).  \label{011}
\end{equation}
To clarify notation we introduce differential operators
\begin{equation}
\overrightarrow{\mathbf{D}^{t}}(\mathbf{R})=\mathbf{1}\left( 1+\frac{a^{2}}{6}\mathbf{\nabla }_{\mathbf{R}}^{2}\right) ,
\quad \overleftarrow{\mathbf{D}^{t}}(\mathbf{R})=\mathbf{1}\left( 1+\frac{a^{2}}{6}\overleftarrow{\mathbf{\nabla }}_{\mathbf{R}}^{2}\right) ,  \label{006}
\end{equation}
\begin{equation}
\left[ \overrightarrow{\mathbf{D}^{r}}(\mathbf{R})\right] _{\alpha \beta }=-\frac{1}{2}\epsilon _{\alpha \beta \gamma }\frac{\partial }{\partial R_{\gamma }},
\quad \left[ \overleftarrow{\mathbf{D}^{r}}(\mathbf{R})\right] _{\alpha \beta }=\frac{1}{2}\epsilon _{\alpha \beta \gamma }\overleftarrow{\frac{\partial }{\partial R_{\gamma }}}.  \label{008}
\end{equation}
where arrow points to the direction of action of differentiation operator.
We rewrite (\ref{eq:v0t}) and (\ref{eq:v0r}) using these operators
\begin{equation}
\mathbf{v}_{0}^{t}(\mathbf{r)}=\mathbf{T}_{0} (\mathbf{r-R}_{j})\cdot \overleftarrow{\mathbf{D}^{t}}(\mathbf{R}_{j})\cdot \bm{\mathcal{F}}, \quad
\mathbf{v}_{0}^{r}(\mathbf{r)}=\mathbf{T}_{0}(\mathbf{r-R}_{j})\cdot \overleftarrow{\mathbf{D}^{r}}(\mathbf{R}_{j})\cdot \bm{\mathcal{T}},\quad|\mathbf{r-R}_{j}|>a.  \label{002}
\end{equation}
For the external flow $\mathbf{v}_{0}(\mathbf{r})$\ which is regular (has no sources within sphere $i$), by the use of
the definition of $\mathbf{w}_{i}^{p}$ (\ref{eq:wtr}), the Faxen laws may be written in analogy to (\ref{eq:Faxen_Laws}) 
\begin{equation}
\mathbf{U}_i = \int_{S_i} \left[ \mathbf{w}^{t}_i (\mathbf{r}') \right]^{T} \cdot \mathbf{v}_0 (\mathbf{r}') d \sigma' = \overrightarrow{\mathbf{D}^{t}}(\mathbf{R}_{i})\cdot \mathbf{v}_0 (\mathbf{R}_i), \quad
\bm{\Omega}_i = \int_{S_i} \left[ \mathbf{w}^{r}_i (\mathbf{r}') \right]^{T} \cdot \mathbf{v}_0 (\mathbf{r}') d \sigma' = \overrightarrow{\mathbf{D}^{r}}(\mathbf{R}_{i})\cdot \mathbf{v}_0 (\mathbf{R}_i).
\label{eq:Faxen_Laws_general}
\end{equation}
We can now write down the Rotne-Prager-Yamakawa mobilities for the unbounded space
(for Oseen propagator $\mathbf{T}_{0})$\ for$\ R_{ij}>2a$ using the
differential operators
\begin{equation}
\bm{\mu }_{ij}^{pq}=\overrightarrow{\mathbf{D}^{p}}(\mathbf{R}_{i})\cdot 
\mathbf{T}_{0}(\mathbf{R}_{i}\mathbf{-R}_{j})\cdot \overleftarrow{\mathbf{D}^{q}}(\mathbf{R}_{j}).  \label{010}
\end{equation}

Now we decompose the arbitrary propagator $\mathbf{T}(\mathbf{r}',\mathbf{r}'')$ as follows
\begin{equation}
\mathbf{T}(\mathbf{r}',\mathbf{r}'')=\left[ \mathbf{T}(\mathbf{r}' 
,\mathbf{r}'')-\mathbf{T}_{0}(\mathbf{r}'-\mathbf{r}'')\right] +
\mathbf{T}_{0}(\mathbf{r}'-\mathbf{r}'')=\mathbf{T}^{\prime }(\mathbf{r}',\mathbf{r}'')+\mathbf{T}_{0}(\mathbf{r}'-\mathbf{r}'').
\label{012}
\end{equation}
The operator $\mathbf{T}'=\mathbf{T}-\mathbf{T}_0$ has no singularities at $\mathbf{r}'=\mathbf{r}''$,
thus see (\ref{eq:Faxen_Laws_general}), it has the property
\begin{equation}
\left\langle \mathbf{w}_{i}^{p}(\mathbf{r}')|\mathbf{T}^{\prime }(\mathbf{r}',\mathbf{r}'')|\mathbf{w}_{j}^{q}(\mathbf{r}'')\right\rangle =
\overrightarrow{\mathbf{D}^{p}}(\mathbf{R}_{i})\cdot \mathbf{T}^{\prime }(\mathbf{R}_{i},\mathbf{R}_{j})\cdot \overleftarrow{\mathbf{D}^{q}}(\mathbf{R}
_{j}).  \label{013}
\end{equation}
Using (\ref{012}) and (\ref{013}) we can cast the mobility $\bm{\mu}_{ij}^{pq}$ in the following form
\begin{eqnarray}
\bm{\mu}_{ij}^{pq} &=&\overrightarrow{\mathbf{D}^{p}}(\mathbf{R}_{i})\cdot \mathbf{T}(\mathbf{R}_{i},\mathbf{R}_{j})\cdot \overleftarrow{\mathbf{D}^{q}}(\mathbf{R}_{j})+\left[ \left\langle \mathbf{w}_{i}^{p}|\mathbf{T}_{0}|\mathbf{w}_{j}^{q}\right\rangle -\overrightarrow{\mathbf{D}^{p}}(\mathbf{R}_{i})\cdot \mathbf{T}_{0}(\mathbf{R}_{i}\mathbf{-R}_{j}) \cdot \overleftarrow{\mathbf{D}^{q}}(\mathbf{R}_{j}) \right] \nonumber \\
&=&\overrightarrow{\mathbf{D}^{p}}(\mathbf{R}_{i})\cdot \mathbf{T}(\mathbf{R}_{i},\mathbf{R}_{j})\cdot \overleftarrow{\mathbf{D}^{q}}(\mathbf{R}_{j})+\mathbf{Y}^{pq}\mathbf{(R}_{ij}).  \label{014}
\end{eqnarray}
The correction
\begin{equation}
\mathbf{Y}^{pq}\mathbf{(R}_{ij}) = \left\langle \mathbf{w}_{i}^{p}|\mathbf{T}_{0}|\mathbf{w}_{j}^{q}\right\rangle -\overrightarrow{\mathbf{D}^{p}}(\mathbf{R}_{i})\cdot \mathbf{T}_{0}(\mathbf{R}_{i}\mathbf{-R}_{j}) \cdot \overleftarrow{\mathbf{D}^{q}}(\mathbf{R}_{j}), \label{eq:Y_correction}
\end{equation}
is non zero only for $|\mathbf{R}_{ij}|<2a$ and is independent of the propagator $\mathbf{T}(\mathbf{r}_{i}\mathbf{,r}_{j})$.
We~write down explicitly the corrections for all components of the mobility matrix (\ref{eq:mutt_final}),(\ref{eq:murr_final}),(\ref{eq:murt_final})
\begin{equation} \label{Y_mutt_correction}
\begin{array}{ccc}  
\mathbf{Y}^{tt}(\mathbf{R}_{ij}) & = & \Theta(2a-R_{ij}) \left\{ \frac{1}{\zeta^{tt}}\left[\left(1-\frac{9R_{ij}}{32a}\right)\mathbf{1}+\frac{3R_{ij}}{32a}\hat{\mathbf{R}}_{ij}\hat{\mathbf{R}}_{ij}\right] \right. \\
& & \left. -\frac{1}{8\pi\eta R_{ij}}\left[\left(1+\frac{2a^{2}}{3R_{ij}^{2}}\right)\mathbf{1}+\left(1-\frac{2a^{2}}{R_{ij}^{2}}\right)\hat{\mathbf{R}}_{ij}\hat{\mathbf{R}}_{ij}\right] \right\},
\end{array}
\end{equation}
\begin{equation} \label{Y_murr_correction}
 \begin{array}{ccc}  
 \mathbf{Y}^{rr}(\mathbf{R}_{ij}) &=& \Theta(2a-R_{ij}) \left\{ \frac{1}{\zeta^{rr}}\left[\left(1-\frac{27}{32}\frac{R_{ij}}{a}+\frac{5}{64}\frac{R_{ij}^{3}}{a^{3}}\right)\mathbf{1}+\left(\frac{9}{32}\frac{R_{ij}}{a}-\frac{3}{64}\frac{R_{ij}^{3}}{a^{3}}\right)\hat{\mathbf{R}}_{ij}\hat{\mathbf{R}}_{ij}\right] \right. \\
 & & \left. + \frac{1}{16\pi\eta R_{ij}^{3}}\left(\mathbf{1}-3\hat{\mathbf{R}}_{ij}\hat{\mathbf{R}}_{ij}\right) \right\},
\end{array}
 \end{equation}
\begin{equation} \label{Y_murt_correction}
 \mathbf{Y}^{rt}(\mathbf{R}_{ij}) =  \mathbf{Y}^{tr}(\mathbf{R}_{ij})
 = \Theta(2a-R_{ij}) \left\{ \frac{1}{16\pi\eta a^{2}}\left(\frac{R_{ij}}{a}-\frac{3}{8}\frac{R_{ij}^{2}}{a^{2}}\right)\boldsymbol{\epsilon}\cdot\hat{\mathbf{R}}_{ij} - \frac{1}{8\pi\eta R_{ij}^{2}}\boldsymbol{\epsilon}\cdot\hat{\mathbf{R}}_{ij} \right\}.
\end{equation}
For the self case, $i=j$ the mobility $\bm{\mu}_{ii}^{pq}$ is obtained
from eq. (\ref{014}, upper line) in the limit $\mathbf{R}_{j}\rightarrow 
\mathbf{R}_{i}$
\begin{eqnarray}
\bm{\mu}_{ii}^{pq} &=&\lim_{\mathbf{R}_{j}\mathbf{\rightarrow R}_{i}}\left\langle \mathbf{w}_{i}^{p}(\mathbf{r}')|\mathbf{T}^{\prime }(\mathbf{r}',\mathbf{r}'')|\mathbf{w}_{j}^{q}(\mathbf{r}'')\right\rangle +\frac{1}{\mathbf{\varsigma }_{0}^{pq}}  \label{015} \\
&=&\lim_{\mathbf{R}_j \rightarrow \mathbf{R}_{i}}\left[ \overrightarrow{\mathbf{D}^{p}}(\mathbf{R}_j)\cdot \mathbf{T}(\mathbf{R}_j,\mathbf{R}_{i})\cdot \overleftarrow{\mathbf{D}^{q}}(\mathbf{R}_{i})-\overrightarrow{\mathbf{D}^{p}}(\mathbf{R}_j)\cdot \mathbf{T}_{0}(\mathbf{R}_j-\mathbf{R}_{i}) \cdot \overleftarrow{\mathbf{D}^{q}}(\mathbf{R}_{i}) \right] +\frac{1}{\mathbf{\varsigma }_{0}^{pq}}.  \nonumber
\end{eqnarray}

To sum up, we have shown how to evaluate the Rotne-Prager-Yamakawa approximation for an arbitrary propagator $\mathbf{T}(\mathbf{r}_{i}\mathbf{,r}_{j})$
by applying to $\mathbf{T}(\mathbf{r}_{i}\mathbf{,r}_{j})$ the differential operators in order to avoid the explicit and often infeasible surface integration.
This allows one to construct the positive definite hydrodynamic
tensors in systems with non-trivial geometry (e.g. in the
presence of a wall, in a channel or in periodic systems). For example, taking in (\ref{014}) the Green's function for a
Stokeslet in the presence of a wall \citep{Blake} leads (for
non-overlapping spheres) to the Rotne-Prager-Blake tensor derived before
by \citet{Bossis1991}, see also \citet{Kim2006,Gauger2009}, and
\citet{Sing2010}. However, these authors did not derive the
regularizing correction for this tensor, which also prevented them
from obtaining the self-term in a manner analogous to our Eq. (\ref{015}). 

On a final note, let us stress that the regularizing correction (\ref{eq:Y_correction})
has the same simple analytical form in all cases, independently of the
particular Green's function $\mathbf{T}(\mathbf{r}_{i}\mathbf{,r}_{j})$.

\section{Concluding remarks}

In this paper, we have re-visited the problem of constructing Rotne-Prager-Yamakawa approximation for mobility and shear disturbance matrices.
A systematic method was presented which allows one to derive the RPY approximation in a systematic way, for translational, rotational and dipolar components of the generalized mobility matrix, both for non-overlapping and overlapping particles.
The regularization corrections for translational-rotational and rotational-rotational mobility tensors have not been previously derived.
These regularizations are crucial in obtaining positive-definite hydrodynamic matrices, which is essential for the Brownian Dynamics simulations.
The positive definiteness also allows for the evaluation of the diffusion tensor and mobility for the bead models (including overlapping beads) of complicated molecules.
Additionally, we have shown how our approach can be generalized to other boundary conditions and corresponding propagators.
\\

EW and KM  acknowledge the support of the Polish National Science Centre (Grant No 2012/05/B/ST8/03010).
PJZ acknowledges support of the Foundation for Polish Science (FNP) through TEAM/2010-6/2 project co-financed by the EU European Regional Development Fund.
PS acknowledges the support of the Polish Ministry of Science and Higher Education (Grant No N N202 055440).

\bibliographystyle{jfm}

\section*{Supporting information: Calculation of integrals for overlapping particles}

The configuration of the two spheres $i$ and $j$ and the notation
is presented in Figure 1. The $z$ axis is chosen in the direction
of the vector $\mathbf{R}_{ij}$ connecting the centres of the spheres.
The variable $\mathbf{r}_{i}$ is a position vector with respect to the centre
of the sphere $i$ and $\mathbf{r}'$ - a position vector with respect to the centre of the sphere $j$. To demonstrate the method it is enough
to calculate explicitly one of the integrals appearing in (3.6)-(3.7) 
and (3.21)-(3.22).
We take the first one, 
\begin{equation}
\mathbf{u}^{t}\left(\mathbf{R}_{ij}\right)=\frac{1}{4\pi a^{2}}\int_{S_{i}}\mathbf{v}_{0}^{t}\left(\mathbf{r}'\right)\mathrm{d}\sigma_{i}.\label{eq:ut-1}
\end{equation}
\begin{figure}
\begin{centering}
\includegraphics[scale=0.25]{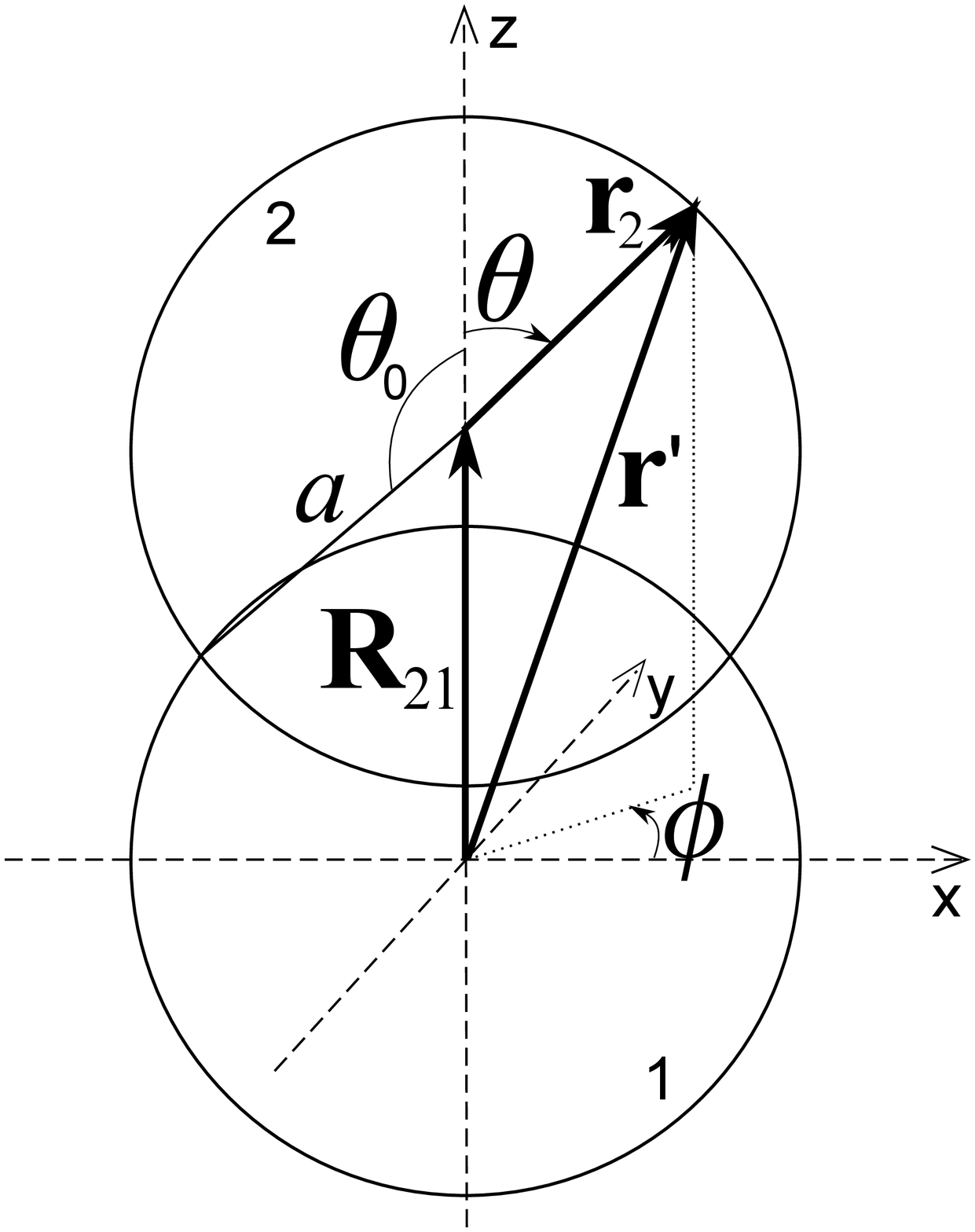}
\par\end{centering}

{\footnotesize{Figure 1. The axes positions and notation for calculation
of the expressions for overlapping spheres.}}
\end{figure}
\noindent The integration is performed over the surface of sphere $i$,
thus $\mathbf{r}_{i}$ is expressed in spherical coordinates $\left(r,\theta,\phi\right)$
associated with sphere $i$ which leads to
\begin{equation}
\bm{\mu}_{ij}^{tt}=\frac{1}{32\pi^{2}\eta}\int_{0}^{\theta_{0}}\mathrm{d}\theta\sin\theta\int_{0}^{2\pi}\mathrm{d}\phi\frac{1}{r'}\left[\left(1+\frac{a^{2}}{3r^{\prime2}}\right)\mathbf{1}+\left(1-\frac{a^{2}}{r^{\prime2}}\right)\hat{\mathbf{r}'}\hat{\mathbf{r}'}\right]+\frac{1}{2\zeta^{tt}}\mathbf{1}\left(1-\frac{R_{ij}}{2a}\right),\label{eq:mu_general_appendix}
\end{equation}
\noindent where $\theta_{0}$ is the meridional angle at which the
two spheres intersect (see Figure 1), defined by
\begin{equation}
\cos\theta_{0}=-\frac{R_{ij}}{2a},\label{eq:theta_0_def}
\end{equation}
\noindent and the vector $\mathbf{r}'=\mathbf{R}_{ij}+\mathbf{r}_{i}$
in the Cartesian basis has the form
\begin{equation}
\mathbf{r}'=a\sin\theta\cos\phi\hat{\mathbf{e}}_{x}+a\sin\theta\sin\phi\hat{\mathbf{e}}_{y}+\left(R_{ij}+a\cos\theta\right)\hat{\mathbf{e}}_{z},\qquad r^{\prime2}=R_{ij}^{2}+2R_{ij}a\cos\theta+a^{2}.\label{eq:r_prime}
\end{equation}
\noindent The last term in (\ref{eq:mu_general_appendix}) results from integration of the $r\leq a$ expression in (3.2)  
from $\theta_0$ to $\pi$. All the azimuthal integrals in (\ref{eq:mu_general_appendix}) are easily calculated to yield
\begin{eqnarray}
\bm{\mu}_{ij}^{tt} & = & \frac{1}{\zeta^{tt}}\mathbf{1}\left[\frac{1}{2}\left(1-\frac{R_{ij}}{2a}\right)+\frac{3}{8}\int_{0}^{\theta_{0}}\mathrm{d}\theta\sin\theta\left(\frac{a}{r'}+\frac{a^{3}}{3r^{\prime3}}\right)+\frac{3}{16}\int_{0}^{\theta_{0}}\mathrm{d}\theta\sin^{3}\theta\left(\frac{a^{3}}{r^{\prime3}}-\frac{a^{5}}{r^{\prime5}}\right)\right]\nonumber \\
 &  & +\frac{3}{16\zeta^{tt}}\hat{\mathbf{R}}_{ij}\hat{\mathbf{R}}_{ij}\int_{0}^{\theta_{0}}\mathrm{d}\theta\sin\theta\left(\frac{a^{3}}{r^{\prime3}}-\frac{a^{5}}{r^{\prime5}}\right)\left[2\left(\frac{R_{ij}}{a}+\cos\theta\right)^{2}-\sin^{2}\theta\right],\label{eq:App_mu_tt}
\end{eqnarray}
\noindent and for the choice of coordinate axes as in Figure 1 we
have
\begin{equation}
\hat{\mathbf{R}}_{ij}\hat{\mathbf{R}}_{ij}=\left[\begin{array}{ccc}
0 & 0 & 0\\
0 & 0 & 0\\
0 & 0 & 1
\end{array}\right].\label{eq:RR}
\end{equation}
\noindent The calculation of $\bm{\mu}_{ij}^{tt}$ is now straightforward.
Since
\begin{equation}
\int_{0}^{\theta_{0}}\mathrm{d}\theta\frac{a\sin\theta}{\left(R_{ij}^{2}+2R_{ij}a\cos\theta+a^{2}\right)^{1/2}}=1,\label{eq:APP_I1}
\end{equation}
\begin{equation}
\int_{0}^{\theta_{0}}\mathrm{d}\theta\frac{a^{3}\sin\theta}{\left(R_{ij}^{2}+2R_{ij}a\cos\theta+a^{2}\right)^{3/2}}=\frac{a}{R_{ij}+a},\label{eq:APP_I2}
\end{equation}
\begin{equation}
\int_{0}^{\theta_{0}}\mathrm{d}\theta\frac{a^{3}\sin^{3}\theta}{\left(R_{ij}^{2}+2R_{ij}a\cos\theta+a^{2}\right)^{3/2}}=\frac{8a-3R_{ij}}{12a},\label{eq:App_I3}
\end{equation}
\begin{equation}
\int_{0}^{\theta_{0}}\mathrm{d}\theta\frac{a^{5}\sin^{3}\theta}{\left(R_{ij}^{2}+2R_{ij}a\cos\theta+a^{2}\right)^{5/2}}=\frac{-R_{ij}^{2}-R_{ij}a+8a^{2}}{12a\left(R_{ij}+a\right)},\label{eq:App_I4}
\end{equation}
\begin{equation}
\int_{0}^{\theta_{0}}\mathrm{d}\theta\frac{a\sin\theta\left(R_{ij}+a\cos\theta\right)^{2}}{\left(R_{ij}^{2}+2R_{ij}a\cos\theta+a^{2}\right)^{3/2}}=\frac{3R_{ij}+4a}{12a},\label{eq:App_I5}
\end{equation}
\begin{equation}
\int_{0}^{\theta_{0}}\mathrm{d}\theta\frac{a^{3}\sin\theta\left(R_{ij}+a\cos\theta\right)^{2}}{\left(R_{ij}^{2}+2R_{ij}a\cos\theta+a^{2}\right)^{5/2}}=\frac{R_{ij}^{2}+R_{ij}a+4a^{2}}{12a\left(R_{ij}+a\right)},\label{eq:App_I6}
\end{equation}
\noindent the $tt$ component of the mobility matrix takes the form
\begin{equation}
\bm{\mu}_{ij}^{tt}=\frac{1}{\zeta^{tt}}\left[\left(1-\frac{9R_{ij}}{32a}\right)\mathbf{1}+\frac{3R_{ij}}{32a}\hat{\mathbf{R}}_{ij}\hat{\mathbf{R}}_{ij}\right],\label{eq:APP_mu_tt_final}
\end{equation}
\noindent as in (3.11). 

\end{document}